\documentclass[12pt]{iopart}
\usepackage{amssymb}
\usepackage{graphicx}
\usepackage{amsthm}
\usepackage{psfrag}
\usepackage{dsfont}

\newtheoremstyle{custom}
{15pt} {15pt} {} {} {\bfseries} {} {.5em} {} \theoremstyle{custom}
\newtheorem{definition}{Definition}

\newtheorem{theorem}[definition]{Theorem}

\begin{document}

\title{The Monotonicity of the Gravitational Entropy Scalar within Quiescent Cosmology}
\author{Philip Threlfall and Susan M Scott}
\address{Centre for Gravitational Physics, \\
College of Physical Sciences, \\
The Australian National University,\\
Canberra ACT 0200\\
Australia} \eads{\mailto{phil.threlfall@anu.edu.au},
\mailto{susan.scott@anu.edu.au}}

\begin{abstract}
In this paper we show that Quiescent Cosmology \cite{Barrow, Goode,
Scott} is consistent with Penrose's Weyl Curvature Hypothesis and
the notion of gravitational entropy \cite{Penrose}. Gravitational
entropy, from a conceptual point of view, acts in an opposite
fashion to the more familiar notion of entropy. A closed system of
gravitating particles will coalesce whereas a collection of gas
particles will tend to diffuse; regarding increasing entropy, these
two scenarios are identical. What has been shown previously
\cite{Goode, Scott} is that gravitational entropy at the initial
singularity predicted by Quiescent Cosmology - the Isotropic Past
Singularity (IPS) - tends to zero. The results from this paper show
that not only is this the case but that gravitational entropy
increases as this singularity evolves.\\

In the first section of this paper we present relevant background
information and motivation. In the second section of this paper we
present the main results of this paper. Our third section contains a
discussion of how this result will inspire future research before we
make concluding remarks in our final section.
\end{abstract}
\maketitle
\section{Background and Motivation}
\subsection{Quiescent Cosmology}
Barrow introduced the world to Quiescent Cosmology in 1978
\cite{Barrow} as an attempt to explain the current large scale
isotropy and homogeneity of the Universe. Quiescent Cosmology
effectively states that the Universe began in a highly ordered state
and has evolved away from its highly regular and smooth beginning
because of gravitational attraction \footnote{This is in contrast to
the ideals of Chaotic Cosmology, made famous by Misner
\cite{Misner}}. This means that the reason we continue to observe
large scale regularity is because we exist in an early stage of
cosmological evolution. In order for Quiescent Cosmology to be
compatible with a Big Bang type singularity, it is necessary that
that singularity is one that is isotropic. This type of initial
isotropic singularity was given a rigorous mathematical definition
in 1985 by Goode and Wainwright \cite{Goode} using a conformal
relationship between two spacetimes. The definition given in this
paper is due to Scott \cite{Scott2} who removed the inherent
technical redundancies of the original definition.\\

Goode and Wainwright based their analysis on the beginning of the
universe, but recently there has been increasing interest in
possible future evolutions of the Universe. H\"ohn and Scott
\cite{Scott} introduced different isotropic and anisotropic
definitions that describe possible future end states of the
Universe. Following Goode and Wainwright they also exploited
conformal relationships between spacetimes.
\subsection{Conformal Structures}
In this paper we primarily deal with isotropic structures and thus
we require the conformal definitions that relate to isotropic
initial and final states of the universe. In order for this paper to
be fully appreciated, however, results pertaining to isotropic
structures will be put in context with anisotropic structures; these
definitions will also be presented in this introduction.\\

The isotropic definitions comprise of the Isotropic Past
Singularity, the Isotropic Future Singularity and the Future
Isotropic Universe. The anisotropic definitions of Quiescent
Cosmology are the Anisotropic Future Endless Universe and the
Anisotropic Future Singularity. Any ancillary definitions that are
needed will also be included.
\begin{definition}[Conformally related metric]
A metric $\mathbf{g}$ is said to be conformally related to a metric
$\mathbf{\tilde{g}}$ on a manifold $\mathcal{M}$ if there exists a
conformal factor $\Omega$ such that
\begin{eqnarray}
\mathbf{g} &=& \Omega^{2}\mathbf{\tilde{g}}\textrm{, where $\Omega$
is a strictly positive function on $\mathcal{M}$.}
\end{eqnarray}
\end{definition}
\begin{definition}[Cosmic time function]
For a space-time $(\mathcal{M},g)$, a cosmic time function is a
function $T$ on the manifold $\mathcal{M}$ which increases along
every future-directed causal curve.
\end{definition}
\subsubsection{Isotropic Definitions}
\begin{figure}[h!]
\begin{center}
\psfrag{A}{ } \psfrag{B}{ }
\psfrag{C}{$\begin{array}{l} \mathbf{g} = \Omega^{2}\left(T\right)\tilde{\mathbf{g}}\\
\Omega\left(0\right) = 0\\
\end{array}$}
\psfrag{D}{$\mathbf{u}$} \psfrag{E}{$\tilde{\mathbf{u}}$}
\psfrag{F}{$T = 0$} \psfrag{G}{$T = 0$} \psfrag{H}{$\begin{array}{l}\textrm{Physical space-time}\\
\left(\mathcal{M}, \mathbf{g}\right)\end{array}$}
\psfrag{I}{$\begin{array}{l}\textrm{Unphysical space-time}\\
\left(\tilde{\mathcal{M}}, \tilde{\mathbf{g}}\right)\end{array}$}
\caption{A pictorial interpretation of an IPS. The fluid flow is
represented by $\mathbf{u}$.\\
\\
}\label{IPS}
\includegraphics[scale = 0.35]{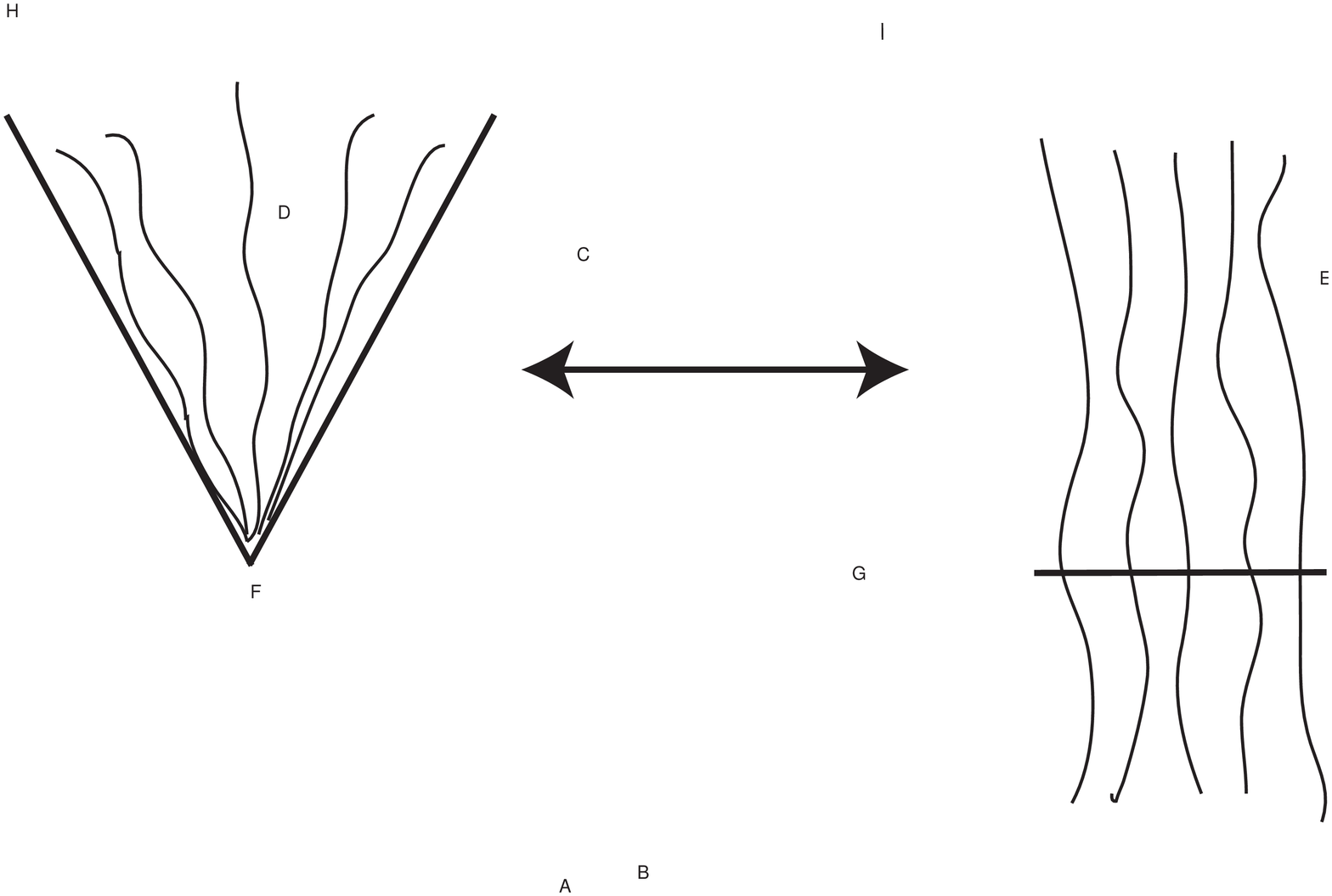}
\end{center}
\end{figure}

It should be noted that we will henceforth denote
relevant quantities for past cosmological frameworks with a tilde
$(\sim)$ and for future cosmological frameworks we will use a bar
$(-)$.
\begin{definition}[Isotropic Past Singularity (IPS)]
A space-time $(\mathcal{M},\mathbf{g})$ admits an Isotropic Past
Singularity if there exists a space-time $(\tilde{\mathcal{M}},
\mathbf{\tilde{g}})$, a smooth cosmic time function $T$ defined on
$\tilde{\mathcal{M}}$ and a conformal factor $\Omega(T)$ which
satisfy
\begin{enumerate}
\item[i)] $\mathcal{M}$ is the open submanifold $T > 0$,
\item[ii)] $\mathbf{g} = \Omega^{2}(T)\mathbf{\tilde{g}}$ on
    $\mathcal{M}$, with $\mathbf{\tilde{g}}$ regular (at least
    $C^{3}$ and non-degenerate) on an open neighbourhood of $T =
    0$,
\item[iii)] $\Omega(0) = 0$ and $\exists\; b > 0$ such that
    $\Omega \in C^{0}[0,b] \cap C^{3}(0,b]$ and $\Omega(0,b] >
    0$,
\item[iv)] $\lambda \equiv \lim\limits_{T\rightarrow 0^{+}}L(T)$
    exists, $\lambda \neq 1$, where $L \equiv
    \frac{\Omega''}{\Omega}\left(\frac{\Omega}{\Omega'}\right)^{2}$
    and a prime denotes differentiation with respect to $T$.
\end{enumerate}\label{IPSDefinition}
\end{definition}
It was demonstrated by Goode and Wainwright \cite{Goode} that, in
order to ensure initial asymptotic isotropy, it is also necessary to
introduce a constraint on the cosmological fluid flow.
\begin{definition}[IPS fluid congruence]
With any unit timelike congruence $\mathbf{u}$ in $\mathcal{M}$ we
can associate a unit timelike congruence $\mathbf{\tilde{u}}$ in
$\tilde{\mathcal{M}}$ such that
\begin{eqnarray}
\mathbf{\tilde{u}} &=& \Omega\mathbf{u}\qquad \textrm{in }
\mathcal{M}\textrm{.}
\end{eqnarray}
\begin{itemize}
\item[a)] If we can choose $\mathbf{\tilde{u}}$ to be regular
    (at least $C^{3}$) on an open neighbourhood of $T = 0$ in
    $\tilde{\mathcal{M}}$, we say that $\mathbf{u}$ is regular
    at the IPS.
\item[b)] If, in addition, $\mathbf{\tilde{u}}$ is orthogonal to
    $T = 0$, we say that $\mathbf{u}$ is orthogonal to the IPS.
\end{itemize}\label{IPSFluidDefinition}
\end{definition}
In figure \ref{IPS} we present a pictorial interpretation of the
IPS.

\begin{figure}[h!]
\begin{center}
 \psfrag{A}{$\bar{T} = 0$} \psfrag{B}{$\bar{T} = 0$}
\psfrag{C}{$\begin{array}{l} \mathbf{g} = \bar{\Omega}^{2}\left(\bar{T}\right)\bar{\mathbf{g}}\\
\bar{\Omega}\left(0\right) = 0\\
\end{array}$}
\psfrag{D}{ } \psfrag{E}{ }
\psfrag{F}{$\begin{array}{l}\textrm{Physical space-time}\\
\left(\mathcal{M}, \mathbf{g}\right)\end{array}$} \psfrag{G}{$\begin{array}{l}\textrm{Unphysical space-time}\\
\left(\bar{\mathcal{M}}, \bar{\mathbf{g}}\right)\end{array}$}
\psfrag{H}{$\mathbf{u}$} \psfrag{I}{$\bar{\mathbf{u}}$} \caption{A
pictorial interpretation of an IFS. The fluid flow is represented by
$\mathbf{u}$. It can be seen that an IFS is essentially a time
reversal of an IPS.\\
\\}
\includegraphics[scale = 0.35]{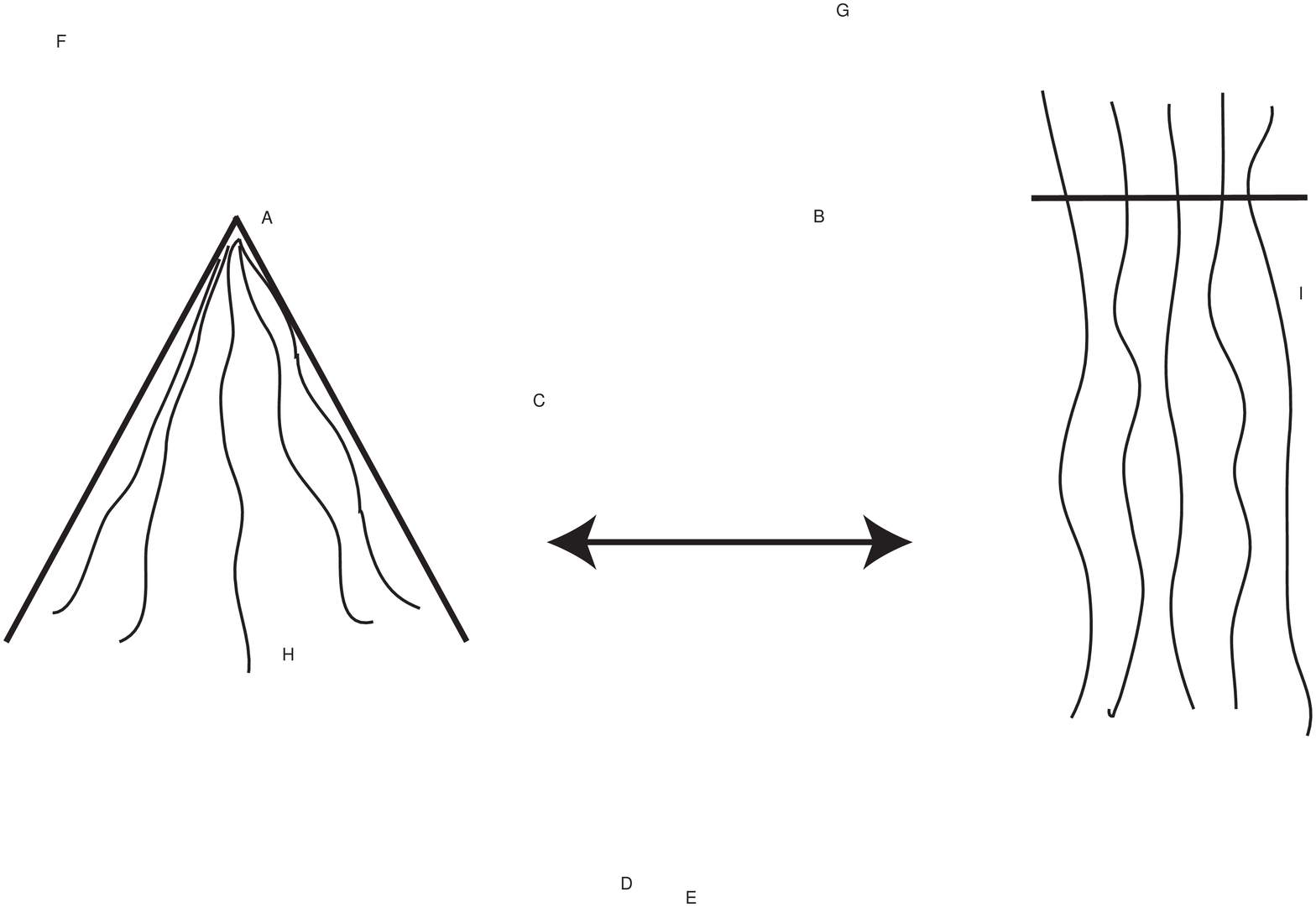}
\label{IFS}
\end{center}
\end{figure}

Below is given the analogous definition of an Isotropic Future
Singularity introduced by H\"ohn and Scott \cite{Scott}, followed by
the constraint on the fluid flow required to ensure final asymptotic
isotropy. The IFS is not compatible with the fundamental ideals of
Quiescent Cosmology (that the end state of the Universe is
anisotropic) but it remains a structure worth analysing.
\begin{definition}[Isotropic Future Singularity (IFS)]
A space-time $(\mathcal{M},\mathbf{g})$ admits an Isotropic Future
Singularity if there exists a space-time $(\bar{\mathcal{M}},
\mathbf{\bar{g}})$, a smooth cosmic time function $\bar{T}$ defined
on $\bar{\mathcal{M}}$, and a conformal factor
$\bar{\Omega}(\bar{T})$ which satisfy
\begin{enumerate}
\item[i)] $\mathcal{M}$ is the open submanifold $\bar{T} < 0$,
\item[ii)] $\mathbf{g} =
    \bar{\Omega}^{2}(\bar{T})\mathbf{\bar{g}}$ on $\mathcal{M}$,
    with $\mathbf{\bar{g}}$ regular (at least $C^{2}$ and
    non-degenerate) on an open neighbourhood of $\bar{T} = 0$,
\item[iii)] $\bar{\Omega}(0) = 0$ and $\exists\; c > 0$ such
    that $\bar{\Omega} \in C^{0}[-c, 0] \cap C^{2}[-c,0)$ and
    $\bar{\Omega}$ is positive on $[-c,0)$,
\item[iv)] $\bar{\lambda} \equiv \lim\limits_{\bar{T}\rightarrow
    0^{-}}\bar{L}(\bar{T})$ exists, $\bar{\lambda} \neq 1$,
    where $\bar{L} \equiv
    \frac{\bar{\Omega}''}{\bar{\Omega}}\left(\frac{\bar{\Omega}}{\bar{\Omega}'}\right)^{2}$
    and a prime denotes differentiation with respect to
    $\bar{T}$.
\end{enumerate}\label{IFSDefinition}
\end{definition}
\begin{definition}[IFS fluid congruence]
With any unit timelike congruence $\mathbf{u}$ in $\mathcal{M}$ we
can associate a unit timelike congruence $\mathbf{\bar{u}}$ in
$\bar{\mathcal{M}}$ such that
\begin{eqnarray}
\mathbf{\bar{u}} &=& \bar{\Omega}\mathbf{u}\qquad \textrm{in }
\mathcal{M}\textrm{.}
\end{eqnarray}
\begin{itemize}
\item[a)] If we can choose $\mathbf{\bar{u}}$ to be regular (at
    least $C^{2}$) on an open neighbourhood of $\bar{T} = 0$ in
    $\bar{\mathcal{M}}$, we say that $\mathbf{u}$ is regular at
    the IFS.
\item[b)] If, in addition, $\mathbf{\bar{u}}$ is orthogonal to
    $\bar{T} = 0$, we say that $\mathbf{u}$ is orthogonal to the
    IFS.
\end{itemize}\label{IFSFluidDefinition}
\end{definition}
In figure \ref{IFS} we present a pictorial interpretation of the
IFS.

Finally we give below the definition of a Future Isotropic Universe
introduced by H\"ohn and Scott \cite{Scott}. This definition covers
the further possibility for a conformal structure with an isotropic
future behaviour, which does not necessarily lead to a future
singularity; for example, some open FRW universes satisfy this
definition \cite{Threlfall}.
\begin{definition}[Future Isotropic Universe
(FIU)] A space-time $(\mathcal{M},\mathbf{g})$ is said to be a
Future Isotropic Universe if there exists a space-time
$(\bar{\mathcal{M}}, \mathbf{\bar{g}})$, a smooth cosmic time
function $\bar{T}$ defined on $\bar{\mathcal{M}}$, and a conformal
factor $\bar{\Omega}(\bar{T})$ which satisfy
\begin{enumerate}
\item[i)] $\lim\limits_{\bar{T}\rightarrow
    0^{-}}\bar{\Omega}(\bar{T}) = +\infty$ and $\exists\; c > 0$
    such that $\bar{\Omega} \in C^{2}[-c,0)$ and $\bar{\Omega}$
    is strictly monotonically increasing and positive on
    $[-c,0)$,
\item[ii)] $\bar{\lambda}$ as defined above exists,
    $\bar{\lambda} \neq 1,2$, and $\bar{L}$ is continuous on
    $[-c,0)$ and
\item[iii)] otherwise the conditions of definitions
    \ref{IFSDefinition} and \ref{IFSFluidDefinition} are
    satisfied.
\end{enumerate}\label{FIUDefinition}
\end{definition}
\subsubsection{Anisotropic Definitions}
All anisotropic definitions\footnote{The definitions given here are
slightly modified from the original ones \cite{Scott}. The
modifications are due to the removal of the limiting causal future
at $\bar{T} = 0$. This has been replaced with our open neighbourhood
on $\bar{T} = 0$.} refer to the future and hence their unphysical
quantities are denoted by a bar ($-$). It is conceivable, however,
that these definitions could be recast for past cosmological states
(similar to the IPS/IFS scenarios).
\begin{definition}[Causal Degeneracy]
Consider $p\in\mathcal{M}$. Let $\gamma_{p}(s)$ be a future
inextendible causal curve such that
$\gamma_{p}(s):[0,a)\to\mathcal{M}$, where
$a\in\mathbb{R}^{+}\cup\{\infty\}$, such that $p =
\gamma_{p}(a)\equiv\mathop{\lim}\limits_{s\to a}\gamma_{p}(s)$ with
limiting tangent vector $\gamma'_{p}\neq0$ at $p$.The metric
$\bar{g}$ is said to be causally degenerate at $p$ if there exists a
curve $\gamma_{p}$ which satisfies $\bar{g}(\gamma'_{p},X) =
0\;\forall\; X\in T_{p}\bar{M}$. (Note that this assumes the metric
is continuous on an open neighbourhood of $p$).
\end{definition}

\begin{figure}[h!]
\begin{center}
 \psfrag{A}{$\bar{T} = 0$} \psfrag{B}{$\bar{T} = 0$}
\psfrag{C}{$\begin{array}{l} \mathbf{g} = \bar{\Omega}^{2}\left(\bar{T}\right)\bar{\mathbf{g}}\\
\bar{\Omega}\left(0\right) = +\infty\\
\mathop {\lim}\limits_{\bar{T} \to
    0^{-}}\bar{\Omega}^{6}|\bar{g}| = +\infty\\
\end{array}$}
\psfrag{F}{$\begin{array}{l}\textrm{Physical spacetime}\\
\left(\mathcal{M}, \mathbf{g}\right)\end{array}$} \psfrag{G}{$\begin{array}{l}\textrm{Unphysical spacetime}\\
\left(\bar{\mathcal{M}}, \bar{\mathbf{g}}\right)\end{array}$}
\psfrag{I}{$\mathbf{u}$} \psfrag{H}{$\bar{\mathbf{u}}$}
\caption{A pictorial interpretation of an AFEU.\\
\\ }
\includegraphics[scale = 0.35]{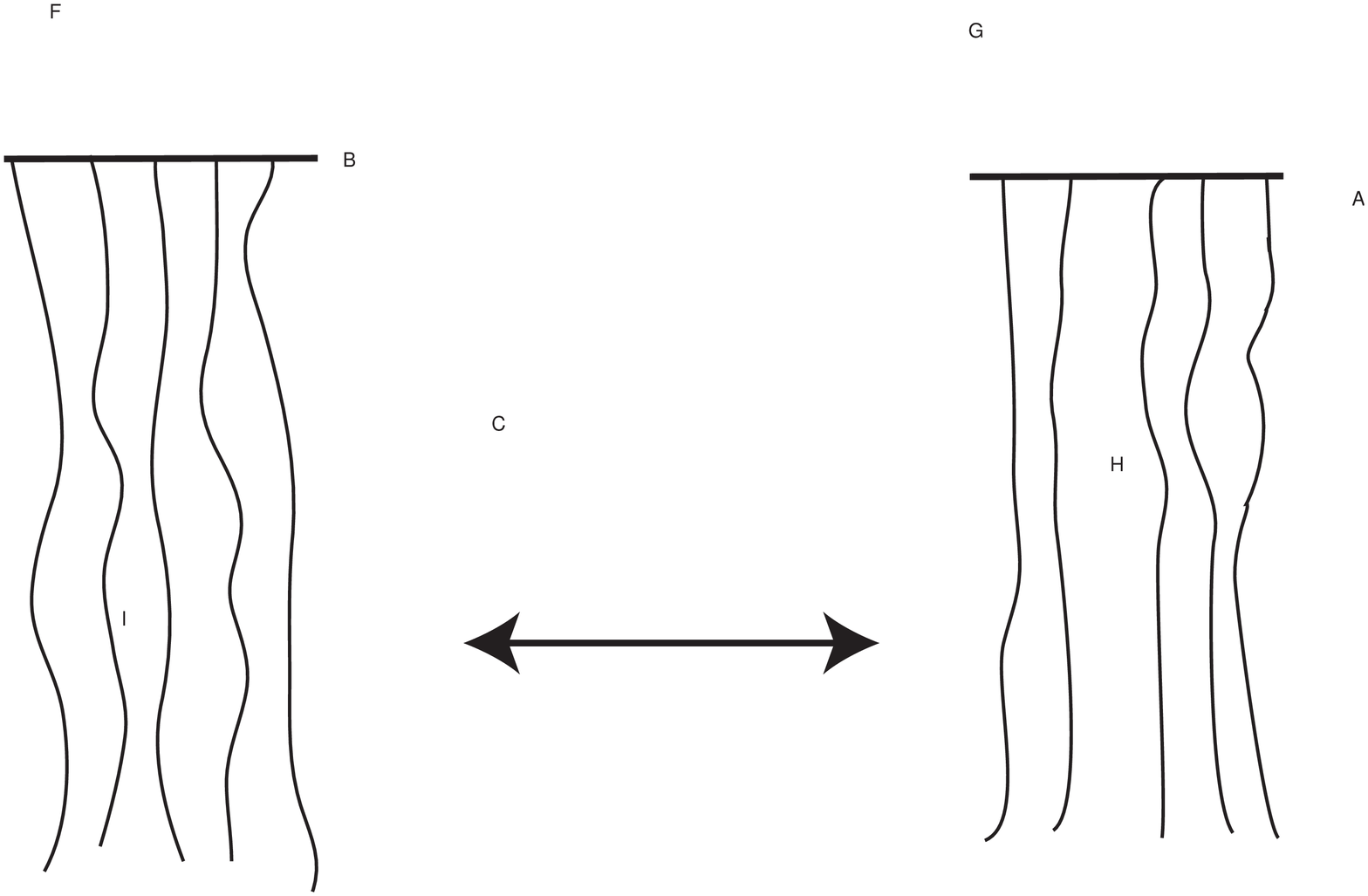}
\label{AFEUDiagram}
\end{center}
\end{figure}

\begin{definition}[Anisotropic Future Endless Universe (AFEU)]
A spacetime, $\left(\mathcal{M}, g\right)$ is said to be an
Anisotropic Future Endless Universe if there exists
\begin{enumerate}
\item a larger manifold $\bar{\mathcal{M}}\supset\mathcal{M}$,
\item a smooth function $\bar{T}$ defined on $\bar{\mathcal{M}}$
    (with $\bar{\nabla}\bar{T}\neq0$ everywhere on
    $\bar{\mathcal{M}})$ such that $\mathcal{M}$ is the open
    submanifold $\bar{T} < 0$,
\item a $C^{0}$ tensor field $\mathbf{\bar{g}}$ of type $(0,2)$
    defined on $\mathcal{M} \cup \mathcal{N}$, where
    $\mathcal{N}$ is an open neighbourhood of $\bar{T} = 0$ in
    $\bar{\mathcal{M}}$, and
\item a conformal factor $\bar{\Omega}({\bar{T}})$ defined on
    $\mathcal{M}$, which satisfies
\begin{enumerate}
\item $\bar{T}$ is a cosmic time function on $\mathcal{M}\:
    \cup\: \mathcal{N}$,
\item $\mathbf{g} =
    \bar{\Omega}^{2}\left(\bar{T}\right)\mathbf{\bar{g}}$ on
    $\mathcal{M}$ and $\mathbf{\bar{g}}$ is degenerate on
    $\bar{T} = 0$,
\item $\mathop{\lim}\limits_{\bar{T} \to
    0^{-}}\bar{\Omega}\left(\bar{T}\right) = +\infty$ and
    $\exists\: c > 0$ such that $\bar{\Omega}\in
    C^{2}[-c,0)$ and $\bar{\Omega}$ is strictly
    monotonically increasing and positive on $[-c,0)$,
\item $\bar{L}$ as defined above is continuous on $[-c,0)$,
    $\bar{\lambda}$ exists, $\bar{\lambda} \neq 1$, and
\item $\mathop {\lim}\limits_{\bar{T} \to
    0^{-}}\bar{\Omega}^{6}|\bar{g}| = \infty $, where
    $\bar{g}$ is the determinant of $\mathbf{\bar{g}}$.
\end{enumerate}
\end{enumerate}
\end{definition}
Thus, the next figure we present is that which represents the AFEU,
seen in figure \ref{AFEUDiagram}.

\begin{figure}[h!]
\begin{center}
 \psfrag{A}{$\bar{T} = 0$} \psfrag{B}{$\bar{T} = 0$}
\psfrag{C}{$\begin{array}{l} \mathbf{g} = \bar{\Omega}^{2}\left(\bar{T}\right)\bar{\mathbf{g}}\\
\bar{\Omega}\left(0\right) = +\infty\\
\mathop {\lim}\limits_{\bar{T} \to
    0^{-}}\bar{\Omega}^{8}|\bar{g}| = 0\\
\end{array}$}
\psfrag{G}{$\begin{array}{l}\textrm{Physical spacetime}\\
\left(\mathcal{M}, \mathbf{g}\right)\end{array}$} \psfrag{F}{$\begin{array}{l}\textrm{Unphysical spacetime}\\
\left(\bar{\mathcal{M}}, \bar{\mathbf{g}}\right)\end{array}$}
\psfrag{H}{$\mathbf{u}$} \psfrag{I}{$\bar{\mathbf{u}}$}
\caption{A
pictorial interpretation of an AFS.\\
\\}
\includegraphics[scale = 0.35]{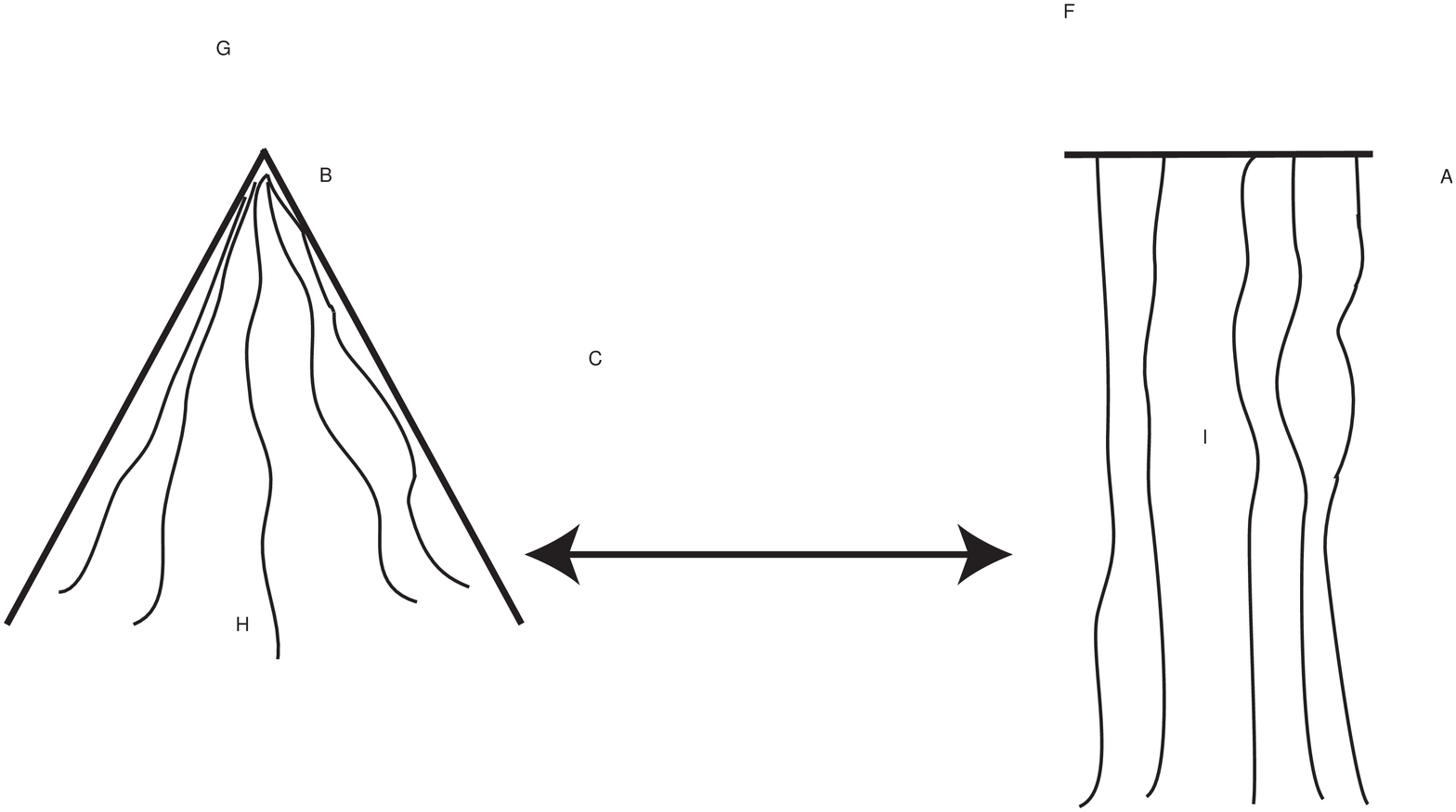}
\label{AFSDiagram}
\end{center}
\end{figure}

\begin{definition}[Anistropic Future Singularity (AFS)]
A spacetime, $\left(\mathcal{M}, g\right)$ is said to be an
Anisotropic Future Singularity if there exists
\begin{enumerate}
\item a larger manifold $\bar{\mathcal{M}}\supset\mathcal{M}$,
\item a smooth function $\bar{T}$ defined on $\bar{\mathcal{M}}$
    (with $\bar{\nabla}\bar{T}\neq0$ everywhere on
    $\bar{\mathcal{M}})$ such that $\mathcal{M}$ is the open
    submanifold $\bar{T} < 0$,
\item a $C^{0}$ tensor field $\mathbf{\bar{g}}$ of type $(0,2)$
    defined on $\mathcal{M}\: \cup\: \mathcal{N}$, where
    $\mathcal{N}$ is an open neighbourhood of $\bar{T} = 0$ in
    $\bar{\mathcal{M}}$, and
\item a conformal factor $\bar{\Omega}({\bar{T}})$ defined on
    $\mathcal{M}$, which satisfies
\begin{enumerate}
\item $\bar{T}$ is a cosmic time function on $\mathcal{M}
    \cup \mathcal{N}$,
\item $\mathbf{g} =
    \bar{\Omega}^{2}\left(\bar{T}\right)\mathbf{\bar{g}}$ on
    $\mathcal{M}$ and $\mathbf{\bar{g}}$ is degenerate on
    $\bar{T} = 0$,
\item $\mathop{\lim}\limits_{\bar{T} \to
    0^{-}}\bar{\Omega}\left(\bar{T}\right) = +\infty$ and
$\exists\: c > 0$ such that $\bar{\Omega}\in C^{2}[-c,0)$
and $\bar{\Omega}$ is strictly monotonically increasing and
positive on $[-c,0)$,
\item $\bar{L}$ as defined above is continuous on $[-c,0)$,
    $\bar{\lambda}$ exists, $\bar{\lambda} \neq 1$, and
\item $\mathop{\lim}\limits_{\bar{T} \to
    0^{-}}\bar{\Omega}^{8}|\bar{g}| = 0$, where $\bar{g}$ is
    the determinant of $\mathbf{\bar{g}}$.
\end{enumerate}
\end{enumerate}
\end{definition}
Figure \ref{AFSDiagram} demonstrates the degenerate nature of the
AFS.
\section{Gravitational Entropy}
The Weyl Curvature Hypothesis (WCH) of Penrose \cite{Penrose} states
that the Weyl curvature at an initial (Big Bang) singularity must be
bounded and has been increasing ever since. Gravitational entropy
\cite{GoodeB, BarrowB, Lim, Scott} is closely linked to the WCH as
will be explained here. A common mental image that comes to mind
when thinking of entropy involves a gas expanding within a chamber -
thus maximising the entropy of the system. If this imagery were
applied to a collection of gravitating particles the particles would
attract one another and end up in a system that seems to have less
entropy than when it started. Penrose \cite{Penrose} addressed this
problem by postulating the gravitational entropy of a system reaches
a maximum when gravitational collapse results in a black hole. This
means that a collection of particles that coalesce are actually
increasing the entropy available and thus gravity can be consistent
with thermodynamics. In General Relativity a quantity that is
hypothesised to be a measure of gravitational entropy \cite{Penrose}
is the ratio between the Weyl and Ricci curvature invariants
\begin{eqnarray}\label{GravEntropy}
K &=& \frac{C^{abcd}C_{abcd}}{R^{ef}R_{ef}}\textrm{.}
\end{eqnarray}
Where it is understood that Weyl curvature describes the curvature
purely due to the gravitational field and the Ricci scalar will
describe the curvature due to matter. In order for entropy to have
been globally increasing (as is expected) from the Big Bang, the
entropy at the Big Bang must have been low, i.e. $K = 0$ at the Big
Bang. The original interpretation for this was that the Weyl scalar
must have been identically zero at the Big Bang but this constraint
was too strict \cite{Tod} as it would have ruled out all
cosmological models apart from the Friedmann-Robertson-Walker (FRW)
solutions. The compromise is that the Weyl scalar must be
asymptotically dominated by the Ricci scalar at the Big Bang. The
IPS has been shown to be consistent with this hypothesis when Goode
and Wainwright \cite{Goode} proved that, as an IPS is approached,
$\mathop{\lim}\limits_{\bar{T}\to 0} K = 0$.
\subsection{The K Theorem}
When H\"ohn and Scott \cite{Scott} expanded the Quiescent Cosmology
framework to consider possible future cosmological states, they were
able to show that Goode and Wainwright's result is able to be
expanded to all isotropic structures; this result was published as
The K Theorem.

\begin{theorem}[The K Theorem]\label{KTheorem}
Let $\mathbf{(\mathcal{M},g)}$ and
$\mathbf{(\bar{\mathcal{M}},\bar{g})}$ be two spacetimes which are
related via the conformal structure $\mathbf{g =
\bar{\Omega}^{2}(\bar{T})\bar{g}}$, where $\bar{T}$ is a smooth
cosmic time function defined on
$\mathbf{(\bar{\mathcal{M}},\bar{g})}$ and $\bar{g}$ is
non-degenerate and at least $C^{2}$ on an open neighbourhood of
$\bar{T}=0$. Let one of the following conditions be true
\begin{enumerate}
\item $\bar{T}\to 0^{-}$, $\mathop {\lim}\limits_{\bar{T} \to
    0^{-}}\bar{\Omega}(0)=\infty$ and $\bar{\Omega}$ is
    positive, $C^{2}$ and strictly increasing on some interval
    $(0,c]$
\item $\bar{T}\to 0^{-}$, $\mathop {\lim}\limits_{\bar{T} \to
    0^{-}}\bar{\Omega}(0)=0$ and $\bar{\Omega}$ is positive,
    $C^{2}$ and strictly decreasing on some interval $[-c,0)$
\item $\bar{T}\to 0^{+}$, $\mathop {\lim}\limits_{\bar{T} \to
    0^{+}}\bar{\Omega}(0)=\infty$ and $\bar{\Omega}$ is
    positive, $C^{2}$ and strictly decreasing on some interval
    $(0,c]$
\item $\bar{T}\to 0^{+}$, $\mathop {\lim}\limits_{\bar{T} \to
    0^{+}}\bar{\Omega}(0)=0$ and $\bar{\Omega}$ is positive,
    $C^{2}$ and strictly increasing
 on some interval $(0,c]$
\end{enumerate}
    and $\bar{\lambda}\neq 1$ then $\mathop {\lim}\limits_{\bar{T} \to 0^{\pm}}
    \frac{C^{abcd}C_{abcd}}{R^{ef}R_{ef}} = 0$.
\end{theorem}

Theorem \ref{KTheorem} is important as it shows that, with a few
assumptions placed upon the conformal factor, all conformally
regular spacetimes give asymptotic isotropic behaviour. This
demonstrates that the Quiescent Cosmology is, at least
asymptotically, consistent with Penrose's ideas about Weyl curvature
at an initial singularity. The other half of Penrose's hypothesis
(that gravitational entropy increases after a Big Bang) is yet to be
answered. It is important to know this answer because if $K$ is
always asymptotically zero according to Quiescent Cosmology then it
means that Quiescent Cosmology is not compatible with Penrose's
ideas in full generality. The most telling indicator of how $K$ may
or may not increase is shown by considering the derivative with
respect to cosmic time. If this value is positive as $\bar{T}$
increases from zero, it means that $K$ will be monotonically
increasing away from the IPS; this is in line with the prediction of
Penrose \cite{Penrose}.

\section{The Derivative of the Gravitational Entropy Scalar}
Recall that $K$ is given by
\begin{eqnarray}
K &=& \frac{C^{abcd}C_{abcd}}{R^{ef}R_{ef}}\\
&=&\frac{\bar{\Omega}^{-4}\bar{C}^{abcd}\bar{C}_{abcd}}{R^{ef}R_{ef}}\textrm{.}
\end{eqnarray}
We can now take the partial derivative of this scalar with respect
to cosmic time to obtain the following
\begin{eqnarray}
K' &=& \frac{(\bar{\Omega}^{-4}\bar{C}^{abcd}\bar{C}_{abcd})_{,m}R^{ef}R_{ef}
- \bar{\Omega}^{-4}(R^{ef}R_{ef})_{,m}\bar{C}^{abcd}\bar{C}_{abcd}
}{\left(R^{ef}R_{ef} \right)^{2}}
\end{eqnarray}
It is clear that we will need to know the derivative of the physical
Ricci scalar and the unphysical Weyl scalar; we present those now.

\subsection{The Unphysical Weyl Scalar's Derivative}
The derivative of the physical Weyl scalar is
\begin{eqnarray}
\left(C^{abcd}C_{abcd}\right)' &=& -4\bar{\Omega}^{-5}\bar{C}^{abcd}\bar{C}_{abcd}\bar{T}_{,m}
 + \bar{\Omega}^{-4}\left(\bar{C}^{abcd}\bar{C}_{abcd}\right)'\textrm{.}
\end{eqnarray}
Where the unphysical Weyl tensor's derivative is
\begin{eqnarray}
(\bar{C}_{abcd})'&=&(\bar{R}_{abcd,m} -\frac{1}{4}\bar{g}^{ij}_{\phantom{ij},m}\left((\bar{g}_{ac}\bar{R}_{idjb} -
\bar{g}_{ad}\bar{R}_{icjb})- (\bar{g}_{bc}\bar{R}_{idja} -
\bar{g}_{bd}\bar{R}_{icja})\right)\nonumber\\
&-&\frac{1}{4}\bar{g}^{ij}((\bar{g}_{ac,m}\bar{R}_{idjb} + \bar{g}_{ac}\bar{R}_{idjb,m} -
\bar{g}_{ad,m}\bar{R}_{icjb}-\bar{g}_{ad}\bar{R}_{icjb,m})\nonumber\\
&-& (\bar{g}_{bc,m}\bar{R}_{idja} +\bar{g}_{bc}\bar{R}_{idja,m} -
\bar{g}_{bd}\bar{R}_{icja}- \bar{g}_{bd,m}\bar{R}_{icja,m}))\nonumber\\
&+& \frac{1}{6}(\bar{g}^{ij}_{\phantom{ij},m}\bar{g}^{kl}\bar{R}_{ikjl} + \bar{g}^{ij}\bar{g}^{kl}_{\phantom{kl},m}\bar{R}_{ikjl}
+ \bar{g}^{ij}\bar{g}^{kl}\bar{R}_{ikjl,m} )\nonumber\\
&\cdot&(\bar{g}_{ac,m}\bar{g}_{db} + \bar{g}_{ac}\bar{g}_{db,m}
-\bar{g}_{ad,m}\bar{g}_{cb} - \bar{g}_{ad}\bar{g}_{cb,m}))
\end{eqnarray}
As such the unphysical Weyl scalar's derivative is
\begin{eqnarray}
(\bar{C}^{abcd}\bar{C}_{abcd})_{,m}&=&
(\bar{g}^{an}_{\phantom{an},m}\bar{g}^{bo}\bar{g}^{cp}\bar{g}^{dq} + \bar{g}^{an}\bar{g}^{bo}_{\phantom{bo},m}\bar{g}^{cp}\bar{g}^{dq}
+ \bar{g}^{an}\bar{g}^{bo}\bar{g}^{cp}_{\phantom{cp},m}\bar{g}^{dq}\nonumber\\
& +& \bar{g}^{an}\bar{g}^{bo}\bar{g}^{cp}\bar{g}^{dq}_{\phantom{dq},m})
(\bar{C}_{abcd,m}\bar{C}_{nopq} + \bar{C}_{abcd}\bar{C}_{nopq,m})
\end{eqnarray}
This is the easier derivative to calculate because of the simple
relationship between the unphysical Weyl scalar and its physical
counterpart.

\subsection{The Physical Ricci Scalar's Derivative} A calculation of the physical
Ricci scalar's derivative is more involved than the Weyl scalar's
but the process is similar enough. With this in mind, first recall
the physical Ricci scalar
\begin{eqnarray}
& &R^{ef}R_{ef}=\bar{\Omega}^{-4}\Big(12\left(\frac{\bar{\Omega}'}{\bar{\Omega}}\right)^{4}
\left(\bar{g}^{ei}\bar{T}_{,e}\bar{T}_{,i}\right)^{2}\left(\bar{L}^{2}-\bar{L} +
1\right)\nonumber\\
&-&
2\left(\frac{\bar{\Omega}'}{\bar{\Omega}}\right)^{3}\left(4\left(2-\bar{L}\right)\bar{g}^{ei}\bar{g}^{fj}\bar{T}_{,e}\bar{T}_{,f}\bar{T}_{:ij}
-2\left(4\bar{L} -
1\right)\bar{g}^{ei}\bar{g}^{fj}\bar{T}_{,e}\bar{T}_{,i}\bar{T}_{:fj}\right)\nonumber\\
&+& \left(\frac{\bar{\Omega}'}{\bar{\Omega}}\right)^{2}\Big(
4\bar{g}^{ei}\bar{g}^{fj}\bar{T}_{:ef}\bar{T}_{:ij} + 8\bar{g}^{ei}\bar{g}^{ei}\left(\bar{T}_{:ei}
\right)^{2} +
4\bar{g}^{ei}\bar{g}^{fj}\left(2-\bar{L}\right)\bar{R}_{ef}\bar{T}_{,i}\bar{T}_{,j}\nonumber\\
&-&2\bar{g}^{ei}\left(1+ \bar{L}\right)\bar{R}\bar{T}_{,e}\bar{T}_{,i}
\Big)\nonumber\\
&-&2\left(\frac{\bar{\Omega}'}{\bar{\Omega}}\right)\left(2\bar{g}^{ei}\bar{g}^{fj}\bar{R}_{ef}\bar{T}_{:ij}
+ \bar{g}^{ei}\bar{R}\bar{T}_{:ei}\right) +
\bar{g}^{ei}\bar{g}^{fj}\bar{R}_{ef}\bar{R}_{ij}\Big)\textrm{.}
\end{eqnarray}
This means that the Ricci scalar's derivative is given by
\begin{eqnarray}
& & \left(R^{ef}R_{ef}\right)_{,m} = \left(\bar{\Omega}^{-4}\right)_{,m}\bar{\Omega}^{4}R^{ef}R_{ef}\nonumber\\
&+& \bar{\Omega}^{-4}\Big(12\left(\frac{\bar{\Omega}'}{\bar{\Omega}}\right)^{4}
\left(\bar{g}^{ei}\bar{T}_{,e}\bar{T}_{,i}\right)^{2}\left(\bar{L}^{2}-\bar{L} +
1\right)\nonumber\\
&-&
2\left(\frac{\bar{\Omega}'}{\bar{\Omega}}\right)^{3}\left(4\left(2-\bar{L}\right)\bar{g}^{ei}\bar{g}^{fj}\bar{T}_{,e}\bar{T}_{,f}\bar{T}_{:ij}
- 2\left(4\bar{L} -
1\right)\bar{g}^{ei}\bar{g}^{fj}\bar{T}_{,e}\bar{T}_{,i}\bar{T}_{:fj}\right)\nonumber\\
&+& \left(\frac{\bar{\Omega}'}{\bar{\Omega}}\right)^{2}\Big(
4\bar{g}^{ei}\bar{g}^{fj}\bar{T}_{:ef}\bar{T}_{:ij} + 8\bar{g}^{ei}\bar{g}^{ei}\left(\bar{T}_{:ei}
\right)^{2} +
4\bar{g}^{ei}\bar{g}^{fj}\left(2-\bar{L}\right)\bar{R}_{ef}\bar{T}_{,i}\bar{T}_{,j}\nonumber\\
&-&2\bar{g}^{ei}\left(1+ \bar{L}\right)\bar{R}\bar{T}_{,e}\bar{T}_{,i}
\Big)\nonumber\\
&-&2\left(\frac{\bar{\Omega}'}{\bar{\Omega}}\right)\left(2\bar{g}^{ei}\bar{g}^{fj}\bar{R}_{ef}\bar{T}_{:ij}
+ \bar{g}^{ei}\bar{R}\bar{T}_{:ei}\right) +
\bar{g}^{ei}\bar{g}^{fj}\bar{R}_{ef}\bar{R}_{ij}\Big)_{,m}\textrm{.}
\end{eqnarray}
This equation is simpler if we analyse it one term at a time. To aid
the reader following along with this derivation, the following two
equations may prove helpful
\begin{eqnarray}
\left(\frac{\bar{\Omega}'}{\bar{\Omega}} \right)^{n}_{\phantom{n},m} &=&
\left(\frac{\bar{\Omega}'}{\bar{\Omega}} \right)^{n+1}\left(\bar{L} - 1\right)\bar{T}_{,m}\\
\bar{L}_{,m} &=& \bar{L}\left(\frac{\bar{\Omega}'}{\bar{\Omega}}\left(1-2\bar{L} \right)
+\frac{\bar{\Omega}'''}{\bar{\Omega}''}\right)\bar{T}_{,m}
\end{eqnarray}
For simplicity's sake, the derivative of the physical Ricci scalar
is given one line at a time. Beginning with the top line,
\begin{eqnarray}
 \left(\bar{\Omega}^{-4}\right)_{,m}\bar{\Omega}^{4}R^{ef}R_{ef} &=& -4\frac{\bar{\Omega}'}{\bar{\Omega}^{5}}\bar{T}_{,m}\bar{\Omega}^{4}R^{ef}R_{ef}
\end{eqnarray}
Now the more interesting lines, beginning with the second
\begin{eqnarray}
& &\bar{\Omega}^{-4}\left(12\left(\frac{\bar{\Omega}'}{\bar{\Omega}}\right)^{4}
\left(\bar{g}^{ei}\bar{T}_{,e}\bar{T}_{,i}\right)^{2}\left(\bar{L}^{2}-\bar{L} +
1\right)\right)_{,m}\nonumber\\
&=& 48\bar{\Omega}^{-4}\Big(\left(\frac{\bar{\Omega}'}{\bar{\Omega}}\right)^{5}\left(\bar{L}-1\right)\left(\bar{L}^{2}-\bar{L} +
1\right)
\left(\bar{g}^{ei}\bar{T}_{,e}\bar{T}_{,i}\right)^{2}\bar{T}_{,m}\nonumber\\
&+& 12\left(\frac{\bar{\Omega}'}{\bar{\Omega}}\right)^{4}\Big(
2\left(\bar{g}^{ei}\bar{T}_{,e}\bar{T}_{,i}\right)\left(\bar{g}^{ei}_{\phantom{ei},m}\bar{T}_{,e}\bar{T}_{,i}
+ \bar{g}^{ei}\left(\bar{T}_{,em}\bar{T}_{,i} + \bar{T}_{,e}\bar{T}_{,im} \right)\right)\nonumber\\
&\cdot&\left(\bar{L}^{2}-\bar{L} +1\right) + \left(2\bar{L}-1\right)\left(\bar{g}^{ei}\bar{T}_{,e}\bar{T}_{,i}\right)^{2}\bar{L}_{,m}\Big)\Big)\textrm{,}
\end{eqnarray}
now the third line
\begin{eqnarray}
& &\bar{\Omega}^{-4}\Big(2\left(\frac{\bar{\Omega}'}{\bar{\Omega}}\right)^{3}\Big(\left(8-4\bar{L}\right)\bar{g}^{ei}\bar{g}^{fj}\bar{T}_{,e}\bar{T}_{,f}\bar{T}_{:ij}\nonumber\\
&-&\left(8\bar{L} +
2\right)\bar{g}^{ei}\bar{g}^{fj}\bar{T}_{,e}\bar{T}_{,i}\bar{T}_{:fj}\Big) \Big)_{,m}\nonumber\\
&=& 6\bar{\Omega}^{-4}\Big(\left(\frac{\bar{\Omega}'}{\bar{\Omega}}\right)^{4}\left(\bar{L}-1\right)\Big(\left(8-4\bar{L}\right)\bar{g}^{ei}\bar{g}^{fj}\bar{T}_{,e}\bar{T}_{,f}\bar{T}_{:ij}\nonumber\\
&-&\left(8\bar{L} +
2\right)\bar{g}^{ei}\bar{g}^{fj}\bar{T}_{,e}\bar{T}_{,i}\bar{T}_{:fj}\Big)\bar{T}_{,m}\nonumber\\
&-& 2\left(\frac{\bar{\Omega}'}{\bar{\Omega}}\right)^{3}\Big(4\bar{g}^{ei}\bar{g}^{fj}\bar{T}_{,e}\bar{T}_{,f}\bar{T}_{:ij}\bar{L}_{,m}\nonumber\\
&+& \left(8-4\bar{L}\right)\Big(
\left(\bar{g}^{ei}_{\phantom{ei},m}\bar{g}^{fj} + \bar{g}^{ei}\bar{g}^{fj}_{\phantom{fj},m}\right)\bar{T}_{,e}\bar{T}_{,f}\bar{T}_{:ij}\nonumber\\
&+&
\bar{g}^{ei}\bar{g}^{fj}\Big(\bar{T}_{,em}\bar{T}_{,f}\bar{T}_{:ij} + \bar{T}_{,e}\bar{T}_{,fm}\bar{T}_{:ij}\nonumber\\
&+& \bar{T}_{,e}\bar{T}_{,f}\bar{T}_{:ij,m}\Big)\Big)- 8\left(\bar{L}_{,m}\bar{g}^{ei}\bar{g}^{fj}\bar{T}_{,e}\bar{T}_{,i}\bar{T}_{:fj}\right)\nonumber\\
&-& \left(8\bar{L} +
2\right)\Big(\left(\bar{g}^{ei}_{\phantom{ei},m}\bar{g}^{fj} + \bar{g}^{ei}\bar{g}^{fj}_{\phantom{fj},m}\right)\bar{T}_{,e}\bar{T}_{,i}\bar{T}_{:fj}\nonumber\\
&-& \bar{g}^{ei}\bar{g}^{fj}\left(\bar{T}_{,em}\bar{T}_{,i}\bar{T}_{:fj}
+ \bar{T}_{,e}\bar{T}_{,im}\bar{T}_{:fj} + \bar{T}_{,e}\bar{T}_{,i}\bar{T}_{:fj,m}\right)\Big)\Big)\Big)\textrm{,}
\end{eqnarray}
the fourth line now
\begin{eqnarray}
& & \bar{\Omega}^{-4}\left(\frac{\bar{\Omega}'}{\bar{\Omega}}\right)^{2}\Big(
4\bar{g}^{ei}\bar{g}^{fj}\bar{T}_{:ef}\bar{T}_{:ij} + 8\left(\bar{g}^{ei}\bar{T}_{:ei}
\right)^{2} +
4\left(2-\bar{L}\right)\bar{g}^{ei}\bar{g}^{fj}\bar{R}_{ef}\bar{T}_{,i}\bar{T}_{,j}\nonumber\\
&-&2\left(1+ \bar{L}\right)\bar{g}^{ei}\bar{R}\bar{T}_{,e}\bar{T}_{,i}
\Big)_{,m}\nonumber\\
&=& \bar{\Omega}^{-4}\Big(\left(\frac{\bar{\Omega}'}{\bar{\Omega}}\right)^{3}\left(\bar{L}-1 \right)\Big(
4\bar{g}^{ei}\bar{g}^{fj}\bar{T}_{:ef}\bar{T}_{:ij} + 8\left(\bar{g}^{ei}\bar{T}_{:ei}
\right)^{2}\nonumber\\
&+&4\left(2-\bar{L}\right)\bar{g}^{ei}\bar{g}^{fj}\bar{R}_{ef}\bar{T}_{,i}\bar{T}_{,j}
-2\left(1+ \bar{L}\right)\bar{g}^{ei}\bar{R}\bar{T}_{,e}\bar{T}_{,i}
\Big)\bar{T}_{,m}\nonumber\\
&+& 4\left(\frac{\bar{\Omega}'}{\bar{\Omega}}\right)^{2}\Big(
\left(\bar{g}^{ei}_{\phantom{ei},m}\bar{g}^{fj} +\bar{g}^{ei}\bar{g}^{fj}_{\phantom{fj},m} \right)\bar{T}_{:ef}\bar{T}_{:ij}
 +
\bar{g}^{ei}\bar{g}^{fj}\left(\bar{T}_{:ef,m}\bar{T}_{:ij} + \bar{T}_{:ef}\bar{T}_{:ij,m}\right)\nonumber\\
&+& 4\bar{g}^{ei}\bar{T}_{:ei}
\left(\bar{g}^{ei}_{\phantom{ei},m}\bar{T}_{:ei} + \bar{g}^{ei}\bar{T}_{:ei,m}\right)-
\left(\bar{L}_{,m}\right)\bar{g}^{ei}\bar{g}^{fj}\bar{R}_{ef}\bar{T}_{,i}\bar{T}_{,j}\nonumber\\
&+&
\left(2-\bar{L}\right)\Big(\left(\bar{g}^{ei}_{\phantom{ei},m}\bar{g}^{fj} + \bar{g}^{ei}\bar{g}^{fj}_{\phantom{fj},m}\right)\bar{R}_{ef}\bar{T}_{,i}\bar{T}_{,j}\nonumber\\
&+&\bar{g}^{ei}\bar{g}^{fj}\Big(\bar{R}_{ef,m}\bar{T}_{,i}\bar{T}_{,j}
+
\bar{R}_{ef}\left(\bar{T}_{,im}\bar{T}_{,j} + \bar{T}_{i}\bar{T}_{,jm}\right)\Big)\Big)\nonumber\\
&-&\frac{1}{2}\Big(\bar{L}_{,m}\bar{g}^{ei}\bar{R}\bar{T}_{,e}\bar{T}_{,i}
+\left(1+ \bar{L}\right)\Big(\bar{g}^{ei}_{\phantom{ei},m}\bar{R}\bar{T}_{,e}\bar{T}_{,i}\nonumber\\
&-&\bar{g}^{ei}\Big(\left(\bar{R}_{,m}\bar{T}_{,e}\bar{T}_{,i}\right)
+\bar{R}\left(\bar{T}_{,em}\bar{T}_{,i} + \bar{T}_{,e}\bar{T}_{,im}\right)\Big)\Big)\Big)\Big)\Big)\textrm{,}
\end{eqnarray}
to the last line
\begin{eqnarray}
& & \bar{\Omega}^{-4}\left(2\left(\frac{\bar{\Omega}'}{\bar{\Omega}}\right)\left(2\bar{g}^{ei}\bar{g}^{fj}\bar{R}_{ef}\bar{T}_{:ij}
+ \bar{g}^{ei}\bar{R}\bar{T}_{:ei}\right) +
\bar{g}^{ei}\bar{g}^{fj}\bar{R}_{ef}\bar{R}_{ij} \right)_{,m}\nonumber\\
&=& \bar{\Omega}^{-4}\Big(2\left(\frac{\bar{\Omega}'}{\bar{\Omega}}\right)^{2}\left(\bar{L}-1 \right)\left(2\bar{g}^{ei}\bar{g}^{fj}\bar{R}_{ef}\bar{T}_{:ij}
+ \bar{g}^{ei}\bar{R}\bar{T}_{:ei}\right)\bar{T}_{,m}\nonumber\\
&+& 2\left(\frac{\bar{\Omega}'}{\bar{\Omega}}\right)
\Big(2\left(\bar{g}^{ei}_{\phantom{ei},m}\bar{g}^{fj} + \bar{g}^{ei}\bar{g}^{fj}_{\phantom{fj},m}\right)\bar{R}_{ef}\bar{T}_{:ij}\nonumber\\
&+& 2\bar{g}^{ei}\bar{g}^{fj}\left(\bar{R}_{ef,m}\bar{T}_{:ij} + \bar{R}_{ef}\bar{T}_{:ij,m}\right)
+ \bar{g}^{ei}_{\phantom{ei},m}\bar{R}\bar{T}_{:ei} + \bar{g}^{ei}\left(\bar{R}_{,m}\bar{T}_{:ei} + \bar{R}\bar{T}_{:ei,m}\right) \Big)\nonumber\\
&+& \left(\bar{g}^{ei}_{\phantom{ei},m}\bar{g}^{fj}
+ \bar{g}^{ei}\bar{g}^{fj}_{\phantom{fj},m}\right)\bar{R}_{ef}\bar{R}_{ij} + \bar{g}^{ei}\bar{g}^{fj}\left(\bar{R}_{ef,m}\bar{R}_{ij} + \bar{R}_{ef}\bar{R}_{ij,m}\right)\Big)\textrm{.}
\end{eqnarray}

\section{Asymptotic Monotonicity of $K$}
Thanks to the work in the last section, we are now in a position to
determine the monotonic behaviour of $K$. It is explicit in the
below theorem but to be clear - we will be dealing with a regular
unphysical metric and hence all unphysical metric components, and
their derivatives, will be well behaved at the hypersurface
$\bar{T}=0$. Furthermore, this means that theorem \ref{KPrime} does
not apply to the AFS and AFEU. It is also important to remember that
for a regular unphysical metric (an IPS/IFS or PIU/FIU)
$\bar{\Omega}$ is $C^{3}$ and as such $\bar{L}_{,m}$ will be well
behaved.

\begin{theorem}[The K-Prime Theorem]\label{KPrime}
Let $\mathbf{(\mathcal{M},g)}$ and
$\mathbf{(\bar{\mathcal{M}},\bar{g})}$ be two spacetimes which are
related via the conformal structure $\mathbf{g =
\bar{\Omega}^{2}(\bar{T})\bar{g}}$, where $\bar{T}$ is a smooth
cosmic time function defined on
$\mathbf{(\bar{\mathcal{M}},\bar{g})}$ and $\bar{g}$ is
non-degenerate and at least $C^{2}$ on an open neighbourhood of
$\bar{T}=0$. If
\begin{eqnarray}
\bar{C}^{abcd}\bar{C}_{abcd}\not\equiv 0
\end{eqnarray}
and one of the following conditions are satisfied
\begin{enumerate}
\item[i)] $\bar{T}\to 0^{+}$, $\mathop {\lim}\limits_{\bar{T}
    \to 0^{+}}\bar{\Omega}=0$ and $\bar{\Omega}$ is positive,
    $C^{3}$ and strictly decreasing on some interval $[-c,0)$,
\item[ii)] $\bar{T}\to 0^{-}$, $\mathop {\lim}\limits_{\bar{T}
    \to 0^{-}}\bar{\Omega}=0$ and $\bar{\Omega}$ is positive,
    $C^{3}$ and strictly decreasing on some interval $[-c,0)$,
\item[iii)] $\bar{T}\to 0^{-}$, $\mathop {\lim}\limits_{\bar{T}
    \to 0^{-}}\bar{\Omega}=+\infty$ and $\bar{\Omega}$ is
    positive, $C^{3}$ and strictly increasing
 on some interval $(0,c]$,
\end{enumerate}
then
\begin{eqnarray*}
 \mathop{\lim}\limits_{\bar{T} \to
0^{\pm}}K'>0\textrm{.}
\end{eqnarray*}
If, however
\begin{enumerate}
\item[iv)] $\bar{T}\to 0^{+}$, $\mathop {\lim}\limits_{\bar{T}
    \to 0^{+}}\bar{\Omega}=+\infty$ and $\bar{\Omega}$ is
    positive, $C^{3}$ and strictly increasing
 on some interval $(0,c]$
\end{enumerate}
then
\begin{eqnarray*}
 \mathop{\lim}\limits_{\bar{T} \to
0^{-}}K'<0\textrm{.}
\end{eqnarray*}
\end{theorem}

\emph{Proof}\\

For subcases i) and ii) the dominant term in the Ricci scalar's
derivative is
\begin{eqnarray}
\left(R^{ef}R_{ef}\right)_{,m}
&\approx& 48\bar{M}^{5}\bar{\Omega}\left(\bar{L}-1\right)\left(\bar{L}^{2}-\bar{L} +
1\right)
\left(\bar{g}^{ei}\bar{T}_{,e}\bar{T}_{,i}\right)^{2}\bar{T}_{,m}
\end{eqnarray}
because this contains the highest power of $\bar{M} :=
\frac{\bar{\Omega}'}{\bar{\Omega}^{2}}$ (which is divergent for
these subcases \cite{Scott}); all other terms are either regular or
bounded.\\

For the subcases iii) and iv) the dominant term of the Ricci
scalar's derivative is
\begin{eqnarray}
\left(R^{ef}R_{ef}\right)_{,m}
&\approx& -48\left(\frac{\bar{\Omega}'}{\bar{\Omega}}\right)^{5}\left(\bar{L}-1\right)\left(\bar{L}^{2}-\bar{L} +
1\right)
\left(\bar{g}^{ei}\bar{T}_{,e}\bar{T}_{,i}\right)^{2}\bar{T}_{,m}
\end{eqnarray}
because it contains the highest power of
$\bar{\Omega}'/\bar{\Omega}$\footnote{the behaviour of this function
has been well described previously \cite{Scott}} and all other terms
will be regular or bounded.\\

While for the Ricci scalar, the dominant term is always going to be
\begin{eqnarray}
R^{ef}R_{ef}&\sim&12\left(\frac{\bar{\Omega}'}{\bar{\Omega}}\right)^{4}
\left(\bar{g}^{ei}\bar{T}_{,e}\bar{T}_{,i}\right)^{2}\left(\bar{L}^{2}-\bar{L}
+ 1\right)\textrm{.}
\end{eqnarray}

Initially we consider cases i) and ii). The entropy scalar's
derivative, in this case is give by,
\begin{eqnarray}
K_{,m} &=&
\frac{-4\bar{\Omega}^{-5}(\bar{C}^{abcd}\bar{C}_{abcd})_{,m}R^{ef}R_{ef}\bar{T}_{,m}
- \bar{\Omega}^{-4}(R^{ef}R_{ef})_{,m}\bar{C}^{abcd}\bar{C}_{abcd}
}{\left(R^{ef}R_{ef} \right)^{2}}\\
&\sim& \frac{-4\bar{\Omega}^{-5}(\bar{C}^{abcd}\bar{C}_{abcd})_{,m}(12\left(\frac{\bar{\Omega}'}{\bar{\Omega}}\right)^{4}
\left(\bar{g}^{ei}\bar{T}_{,e}\bar{T}_{,i}\right)^{2}\left(\bar{L}^{2}-\bar{L}
+ 1\right))\bar{T}_{,m}}{\left(12\left(\frac{\bar{\Omega}'}{\bar{\Omega}}\right)^{4}
\left(\bar{g}^{ei}\bar{T}_{,e}\bar{T}_{,i}\right)^{2}\left(\bar{L}^{2}-\bar{L}
+ 1\right)\right)^{2}}\nonumber\\
&-& \frac{\bar{\Omega}^{-4}\left(48\bar{M}^{5}\bar{\Omega}\left(\bar{L}-1\right)\left(\bar{L}^{2}-\bar{L} +
1\right)
\left(\bar{g}^{ei}\bar{T}_{,e}\bar{T}_{,i}\right)^{2}\bar{T}_{,m}\right)\bar{C}^{abcd}\bar{C}_{abcd}
}{\left(12\left(\frac{\bar{\Omega}'}{\bar{\Omega}}\right)^{4}
\left(\bar{g}^{ei}\bar{T}_{,e}\bar{T}_{,i}\right)^{2}\left(\bar{L}^{2}-\bar{L}
+ 1\right)\right)^{2}}\\
&=& \frac{-(\bar{C}^{abcd}\bar{C}_{abcd})_{,m}\bar{T}_{,m}}
{3\bar{\Omega}^{5}\left(\frac{\bar{\Omega}'}{\bar{\Omega}}\right)^{4}
\left(\bar{g}^{ei}\bar{T}_{,e}\bar{T}_{,i}\right)^{2}\left(\bar{L}^{2}-\bar{L}
+ 1\right)}\nonumber\\
&-& \frac{\left(\bar{L}-1\right)\bar{C}^{abcd}\bar{C}_{abcd}\bar{T}_{,m}}
{3\bar{\Omega}^{7}\left(\frac{\bar{\Omega}'}{\bar{\Omega}}\right)^{3}
\left(\bar{g}^{ei}\bar{T}_{,e}\bar{T}_{,i}\right)^{2}\left(\bar{L}^{2}-\bar{L}
+ 1\right)}\\
&\sim& \frac{\left(1-\bar{L}\right)\bar{C}^{abcd}\bar{C}_{abcd}\bar{T}_{,m}}
{3\bar{\Omega}^{7}\left(\frac{\bar{\Omega}'}{\bar{\Omega}}\right)^{3}
\left(\bar{g}^{ei}\bar{T}_{,e}\bar{T}_{,i}\right)^{2}\left(\bar{L}^{2}-\bar{L}
+ 1\right)}\\
&=& \frac{1}{\bar{\Omega}}\frac{1}{\frac{\bar{\Omega}'}{\bar{\Omega}}}
\frac{\bar{C}^{abcd}\bar{C}_{abcd}\left(1-\bar{L}\right)\bar{T}_{,m}}
{3\bar{\Omega}^{6}\left(\frac{\bar{\Omega}'}{\bar{\Omega}}\right)^{2}
\left(\bar{g}^{ei}\bar{T}_{,e}\bar{T}_{,i}\right)^{2}\left(\bar{L}^{2}-\bar{L}
+ 1\right)}\textrm{.}
\end{eqnarray}
The reason for writing it in this form becomes clear when we note
that $\mathop{\lim}\limits_{\bar{T} \to
0^{\pm}}\bar{\Omega}(\bar{T})\to 0$, $\bar{\lambda} < 1$. Therefore
the sign of $K_{,m}$ solely depends on the signs of $\bar{\Omega}$
and
$\bar{\Omega}'/\bar{\Omega}$ because all other terms are positive and nonzero.\\

The reader will recall that for case i), $\mathop
{\lim}\limits_{\bar{T}
    \to 0^{+}}1/\bar{\Omega}\to
\infty^{+}$ and $\mathop {\lim}\limits_{\bar{T}
    \to 0^{+}}\bar{\Omega}'/\bar{\Omega}\to +\infty$. Hence the
above is
always positive and so is the entropy scalar's derivative.\\

For case ii), $\mathop {\lim}\limits_{\bar{T}
    \to 0^{-}}1/\bar{\Omega}\to \infty^{-}$ and
$\mathop {\lim}\limits_{\bar{T}
    \to 0^{-}}\bar{\Omega}'/\bar{\Omega}\to -\infty$
 and hence the above is positive  and so is $K_{,m}$.\\

We turn to cases iii) and iv) now. The entropy scalar's derivative
is
\begin{eqnarray}
& &K_{,m} =
\frac{-4\bar{\Omega}^{-5}(\bar{C}^{abcd}\bar{C}_{abcd})_{,m}R^{ef}R_{ef}\bar{T}_{,m}
- \bar{\Omega}^{-4}(R^{ef}R_{ef})_{,m}\bar{C}^{abcd}\bar{C}_{abcd}
}{\left(R^{ef}R_{ef} \right)^{2}}\\
&\sim& \frac{-4\bar{\Omega}^{-5}(\bar{C}^{abcd}\bar{C}_{abcd})_{,m}(12\left(\frac{\bar{\Omega}'}{\bar{\Omega}}\right)^{4}
\left(\bar{g}^{ei}\bar{T}_{,e}\bar{T}_{,i}\right)^{2}\left(\bar{L}^{2}-\bar{L}
+ 1\right))\bar{T}_{,m}}{\left(12\left(\frac{\bar{\Omega}'}{\bar{\Omega}}\right)^{4}
\left(\bar{g}^{ei}\bar{T}_{,e}\bar{T}_{,i}\right)^{2}\left(\bar{L}^{2}-\bar{L}
+ 1\right)\right)^{2}}\nonumber\\
&+& \frac{\bar{\Omega}^{-4}\left(48\left(\frac{\bar{\Omega}'}{\bar{\Omega}}\right)^{5}\left(\bar{L}-1\right)\left(\bar{L}^{2}-\bar{L} +
1\right)
\left(\bar{g}^{ei}\bar{T}_{,e}\bar{T}_{,i}\right)^{2}\bar{T}_{,m}\right)\bar{C}^{abcd}\bar{C}_{abcd}
}{\left(12\left(\frac{\bar{\Omega}'}{\bar{\Omega}}\right)^{4}
\left(\bar{g}^{ei}\bar{T}_{,e}\bar{T}_{,i}\right)^{2}\left(\bar{L}^{2}-\bar{L}
+ 1\right)\right)^{2}}\\
&=& \frac{-\bar{\Omega}^{-5}(\bar{C}^{abcd}\bar{C}_{abcd})_{,m}\bar{T}_{,m}}{3\left(\frac{\bar{\Omega}'}{\bar{\Omega}}\right)^{4}
\left(\bar{g}^{ei}\bar{T}_{,e}\bar{T}_{,i}\right)^{2}\left(\bar{L}^{2}-\bar{L}
+ 1\right)}\nonumber\\
&+& \frac{\bar{\Omega}^{-4}\frac{\bar{\Omega}'}{\bar{\Omega}}\bar{C}^{abcd}\bar{C}_{abcd}\bar{T}_{,m}
}{3\left(\frac{\bar{\Omega}'}{\bar{\Omega}}\right)^{4}
\left(\bar{g}^{ei}\bar{T}_{,e}\bar{T}_{,i}\right)^{2}\left(\bar{L}^{2}-\bar{L}
+ 1\right)}\\
&\sim&\frac{\bar{C}^{abcd}\bar{C}_{abcd}\bar{T}_{,m}
}{3\bar{\Omega}^{4}\left(\frac{\bar{\Omega}'}{\bar{\Omega}}\right)^{3}
\left(\bar{g}^{ei}\bar{T}_{,e}\bar{T}_{,i}\right)^{2}\left(\bar{L}^{2}-\bar{L}+ 1\right)}
\end{eqnarray}
As we saw before, this mathematical form is helpful because
$\mathop{\lim}\limits_{\bar{T} \to
0^{\pm}}\bar{\Omega}^{4}(\bar{T})\to +\infty$, $\bar{\lambda} > 1$.
So the sign of $K_{,m}$ solely depends on
the sign of $\bar{\Omega}'/\bar{\Omega}$ because all other terms are positive and nonzero.\\

For case iii) $\mathop {\lim}\limits_{\bar{T}
    \to 0^{+}}\bar{\Omega}'/\bar{\Omega}\to -\infty$ and hence the above is negative.\\

For case iv) $\mathop {\lim}\limits_{\bar{T}
    \to 0^{-}}\bar{\Omega}'/\bar{\Omega}\to +\infty$
and hence the above is positive. $\Box$\\

This now completes the proof. In order to guide the reader in
visualising this behaviour, we present three representations of the
monotonicity of $K$. The first represents cases i) and iii), the
second is case ii) and the last is case iv). \newpage
\begin{figure}[h]
\begin{center}
\caption{A representation of how $K$ would behave if it were monotonically increasing away from $\bar{T} = 0$}
\includegraphics[width=0.45\textwidth]{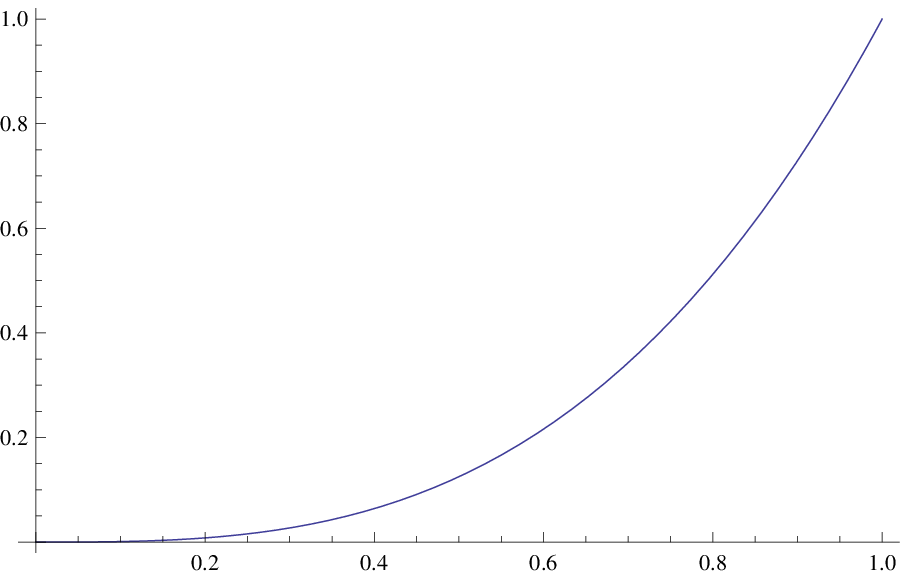}
\label{IPSPIUKPrime}
\caption{A representation of how $K$ would behave if it were monotonically decreasing toward $\bar{T} = 0$}
\includegraphics[width=0.45\textwidth]{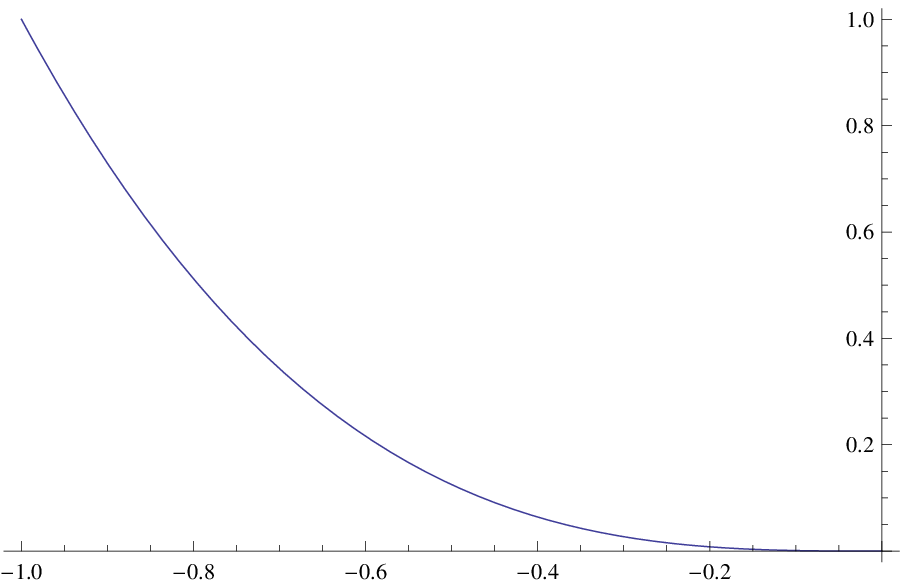}
\label{IFSKPrime}
\caption{A representation of how $K$ would behave if it were monotonically increasing toward $\bar{T} = 0$}
\includegraphics[width=0.45\textwidth]{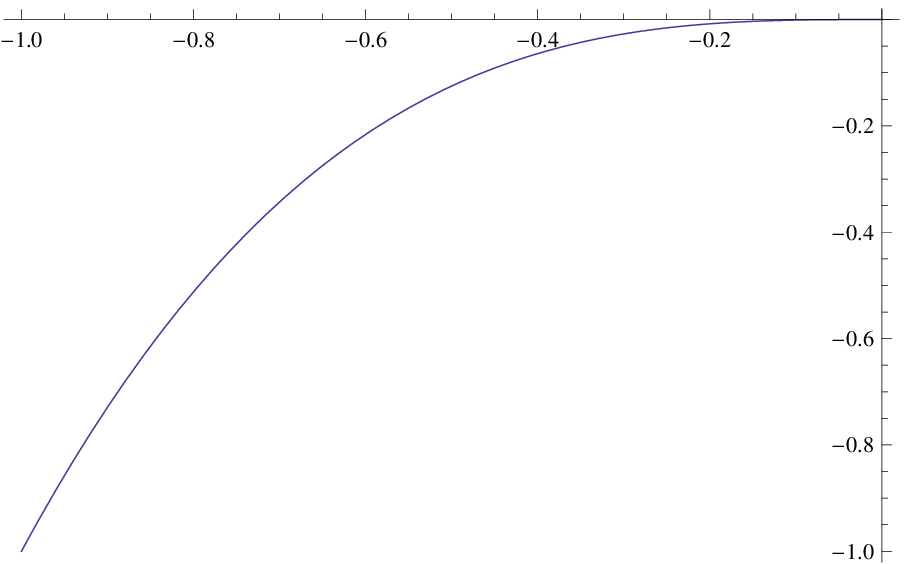}
\label{FIUKPrime}
\end{center}
\end{figure}
This is fundamentally important because it means for initial
isotropic structures, their measure of gravitational entropy will
increase away from zero and for final isotropic states, their
gravitational entropy will decrease towards zero. What we want to
ascertain now is how $K$ behaves for anisotropic future states. At
this stage, all example cosmologies that admit an AFEU or AFS have
$K > 0$ but we have not been able to prove this in all generality.
As the direction of this study will be somewhat different to this
paper, we defer this discussion to an upcoming paper apart from the
following remarks.\\

All observational evidence indicates that, at least locally, entropy
is ever increasing and if Quiescent Cosmology is to be consistent
with observational evidence (as well as Penrose's WCH) then the AFEU
and AFS should have a measure of gravitational entropy that is
nonzero. This will serve to demonstrate that a universe that begins
with an isotropic structure and ends in an anisotropic state will
have a net increase of gravitational entropy. This will not serve to
demonstrate monotonicity in the intermediate region as gravitational
entropy may be oscillatory in nature during this region but it will
show a net increase.
\section{Conclusions and Further Outlook}
The work in this paper is pivotal to prove not only the physical
plausibility of an IPS but also to demonstrate that the IPS is truly
compatible with the WCH. We have been able to show that the
gravitational entropy scalar will, in a local neighbourhood of the
IPS at $\bar{T} = 0$, monotonically increase for non conformally
flat spacetimes. This is in direct agreement with Penrose's
conjecture regarding the dominance of the
Weyl scalar.\\

Furthermore, we have also been able to prove that, if the Universe
did not start with a Big Bang but rather was a uniform distribution
of matter, corresponding to a PIU then this too has zero
gravitational entropy that monotonically increases as cosmic time
increases. Although classical General Relativity seems to predict
that the Universe started with a Big Bang, it is reassuring
nevertheless, that Quiescent Cosmology and the WCH is compatible
with this structure.\\

If the Universe were to end in an isotropic singularity then the
gravitational entropy will be locally monotonically increasing
towards zero. This seems to indicate that $K$ would obtain a maximum
(negative) value before increasing to zero. This is somewhat similar
to the case when the Universe ends as an FIU because in this case
the entropy scalar decreases monotonically as the FIU is approached.
Both of these scenarios indicate that at some stage prior to the
isotropic end, the Universe had a maximum, nonzero value of
gravitational entropy and that it will tend to zero in the future.
This is not compatible with the second law of thermodynamics but it
means that if the end of the Universe is going to be isotropic then
it means gravitational
entropy will have to decrease from some finite maximal value.\\

As mentioned at the end of our main results section, the obvious
extension for this type of work is to consider our anisotropic
futures and see if their gravitational entropy scalar's are
monotonically increasing as they are approached. As will be seen in
future papers, the problems caused by the degenerate nature of the
AFEU and AFS will force us to address the question of gravitational
entropy in a different manner.
\section*{References}
\bibliographystyle{unsrt}
\bibliography{Bibliography}
\end{document}